# Stretched exponential spin relaxation in organic superconductors

Joseph Gezo, Tak-Kei Lui, Brian Wolin, Charles P. Slichter, Russell Giannetta

*Loomis Laboratory of Physics, University of Illinois at Urbana-Champaign, Urbana, Illinois 61801*

John A. Schlueter

*Materials Science Division, Argonne National Laboratory, Argonne, Illinois 60439*

## Abstract

Proton NMR measurements on the organic superconductor κ-(ET)$_2$Cu[N(CN)$_2$]Br $\left(T_C = 11.6\,K\right)$ exhibit stretched exponential spin-lattice relaxation below $T \approx 25\,K$, suggestive of an inhomogeneous magnetic phase that develops in the normal state and coexists with superconductivity. The onset of this phase coincides approximately with a large normal state Nernst signal reported previously. By contrast, the closely related superconductor κ-(ET)$_2$Cu(NCS)$_2$ $\left(T_C = 10.5\,K\right)$ shows single exponential spin-lattice relaxation and a conventional Nernst effect. The temperature range $T_C < T < 30$ K encompasses several phenomena in the κ-(ET)$_2$X conductors, including changes in conduction electron spin resonance, electronic phase separation and the onset of antiferromagnetic order. Analogous behavior in La$_{2-x}$Sr$_x$CuO$_4$ suggests that a density wave may develop in κ-(ET)$_2$Cu[N(CN)$_2$]Br.

The observation of a large normal state Nernst signal in κ-(ET)$_2$Cu[N(CN)$_2$]Br [1] reinforces the correspondence between κ-(ET)$_2$X [2] organic superconductors and the copper oxides [3,4]. In the cuprates, large Nernst signals were first attributed to normal state vortices [5,6] but this interpretation has been challenged by models involving enhanced Gaussian fluctuations [7] and Fermi surface reconstruction [8, 9]. If magnetic fluctuations of some kind are responsible for the Nernst signal they may be detectable via nuclear spin relaxation. With this motivation in mind, we have examined proton NMR in κ-(ET)$_2$Cu[N(CN)$_2$]Br $(T_C = 11.6\,K)$ and κ-(ET)$_2$Cu(NCS)$_2$ $(T_C = 10.5\,K)$. Since ethylene group protons in the ET molecule are weakly coupled to conduction electrons, they are sensitive probes of vortex motion and in principle, other sources of non-hyperfine spin relaxation [10-13]. The separation of hyperfine relaxation from other mechanisms is more difficult to achieve in the copper oxides [14]. We find that below $T \sim 25$ K the spin lattice relaxation in κ-(ET)$_2$Cu[N(CN)$_2$]Br develops a stretched exponential time dependence whose exponent $\beta$ exhibits no apparent change at $T_C$. The temperature and field dependence of $\beta$ shows a striking correspondence to the Nernst coefficient reported earlier [1]. By contrast, spin relaxation in κ-(ET)$_2$Cu(NCS)$_2$, which exhibits no anomalous Nernst effect, shows single exponential time dependence. The temperature range $T_C < T < 30$ K encompasses several phenomena observed in the κ-(ET)$_2$X family, including changes in conduction electron spin resonance [15], electronic phase separation [16,17] and the onset of antiferromagnetic order κ-(ET)$_2$Cu[N(CN)$_2$]Cl at $T_{Neel} = 27$ K [18]. We discuss similarities to La$_{2-x}$Sr$_x$CuO$_4$ where experiments have also demonstrated stretched exponential relaxation [19] and a Nernst signal far above $T_C$ [5,6].

Single crystals of κ-(ET)$_2$Cu[N(CN)$_2$]Br and κ-(ET)$_2$Cu(NCS)$_2$ with natural isotopic abundances were grown using methods previously described [20,21]. The samples were mounted on sapphire with a small amount of proton-free grease [22] and cooled in a gas flow cryostat. It is well established that the rate of cooling through 80 K has a significant effect on the transport and superconducting properties of κ-(ET)$_2$Cu[N(CN)$_2$]Br [23-27]. To ensure a controlled experiment we followed the same cooling procedure for all runs. Except where indicated, the sample was cooled at $dT/dt = 0.3$ K/min from 100 K down to 40 K. As we show later, a factor of 60 change in the cooling rate made no significant difference to our main result. NMR measurements were taken with a homebuilt probe and spectrometer in fields of 1, 1.5, 2 and 3 Tesla. Unless otherwise indicated, the static field $B_0$ was perpendicular to the conducting planes and the RF ($H_1$) field was parallel to the conducting planes. Due to the very weak conduction between planes, the RF skin depth in this orientation is much larger than the sample size. Nutation curves for spin rotation angles out to 450 degrees were weakly decaying sinusoids, indicative of uniform spin excitation throughout the sample. A saturating comb, solid echo sequence, $[90_x]^{50} - t - 90_x - \tau_e - 90_y - \tau_e - solid\ echo$, was used to measure spin lattice relaxation. The $90_x$ comb ($10^{-4}$ sec pulse spacing) ensures uniform saturation. The solid echo ($\tau_e = 10^{-5}$ sec) refocuses nearest neighbor dipolar coupling which is large (~ 40 kHz) for ethylene protons [28].

Figure 1 shows the magnetization recovery versus delay time for κ-(ET)$_2$Cu[N(CN)$_2$]Br at 9 K and $B_0$ = 1 T. Previous investigators have fit recoveries to a sum of two exponentials, the longer of which was considered intrinsic to the sample [10,12]. However, we find that a stretched exponential (solid curve),

$$(1) \quad M(t) = M_0 \left(1 - \exp\left[-(t/T_1)^\beta\right]\right)$$

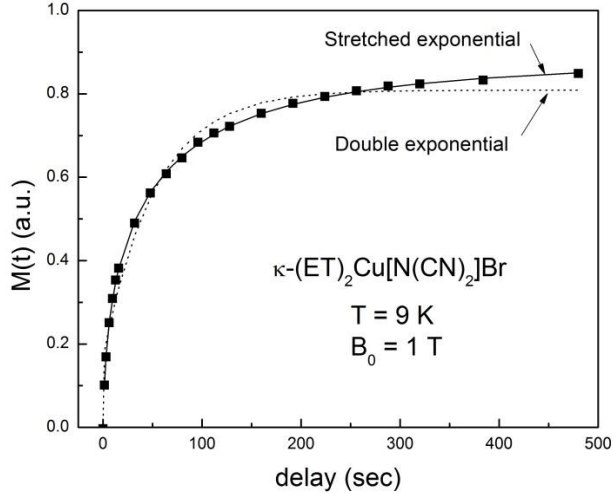

**Fig. 1.** Magnetization recovery versus time after a saturating comb. (Dashed curve) fit to double exponential recovery. (Solid curve) fit to stretched exponential recovery with $T_1$ = 44 sec, $\beta$ = 0.52.

provides a better fit to our data over the entire recovery period. Stretched exponential behavior is routinely observed in a wide variety of disordered systems [29,30]. In the context of NMR it is often viewed as the response to a distribution of spin-lattice relaxation times [19,31]. Small $\beta$ implies a broad distribution while $\beta$ = 1 corresponds to the single exponential relaxation expected for a homogeneous system.

Fig. 2 shows $T_1$ and $\beta$ in κ-(ET)$_2$Cu[N(CN)$_2$]Br for *4.2 K < T < 300 K*, in a field of 1 T. For temperatures above *25 K* the stretching exponent $\beta$ approaches unity, indicating uniform spin lattice relaxation. In this region the double exponential fit is nearly indistinguishable from the stretched exponential and our values for $T_1$ agree with previous measurements [10,11,12,23]. Beginning at *T* = 25 K $\beta$ begins to fall, reaching a minimum of $\beta$ ~ 0.5. Below 7.5 K magnetic fluctuations from the vortex liquid lead to more homogeneous spin relaxation and an increasing $\beta$. For still lower temperatures $\beta$ decreases as the vortex lattice forms and relaxation once again becomes inhomogeneous.

$1/T_1$ exhibits several different regimes of relaxation., shown in Fig. 2. The large peak near 250 K appears as the exponentially-activated correlation time for ethylene group motion crosses the Larmor frequency [10,13,33]. The smaller peak near 160 K coincides with large changes in central $^{13}$C and $^1$H linewidths [16,34,35]. The weak maximum near 50 K coincides with a sharp change in $1/T_1$ for $^{13}$C nuclei that is associated with the opening of a spin gap [34-37]. Nearly all transport coefficients [3] as well as the conduction electron spin resonance signal change dramatically near 50 K [15,38,39]. The diverse physical mechanisms responsible for these changes in $1/T_1$ have no apparent effect on $\beta$. The relaxation remains homogeneous down to $T = 25$ K, below which $\beta$ drops and the slope of $1/T_1$ versus T shows a slight change. Fig. 3 shows corresponding data for κ-(ET)$_2$Cu(NCS)$_2$. In this case single exponential ($\beta \sim 1$) relaxation holds throughout the entire temperature range. The large peak near 250 K and the weaker peak near 50 K remain. The peak near 160 K is less well defined nor is there a vortex peak. The latter is consistent with a much lower irreversibility temperature in this material [40]. Neither material showed any change in spin-lattice relaxation at $T_C$, consistent with a very weak hyperfine coupling to the conduction electron system. In both cases, $1/T_1$ was independent of position in the NMR line. Despite different anions and crystal structures $\kappa$-$(ET)_2Cu[N(CN)_2]Br$ (orthorhombic) and $\kappa$-$(ET)_2Cu(NCS)_2$ (monoclinic) have very similar superconducting properties [41,42] making it unlikely that the unusual spin relaxation seen only in $\kappa$-$(ET)_2Cu[N(CN)_2]Br$ is a superconducting fluctuation effect [43]. In addition, fluctuations would not be expected to produce stretched exponential relaxation.

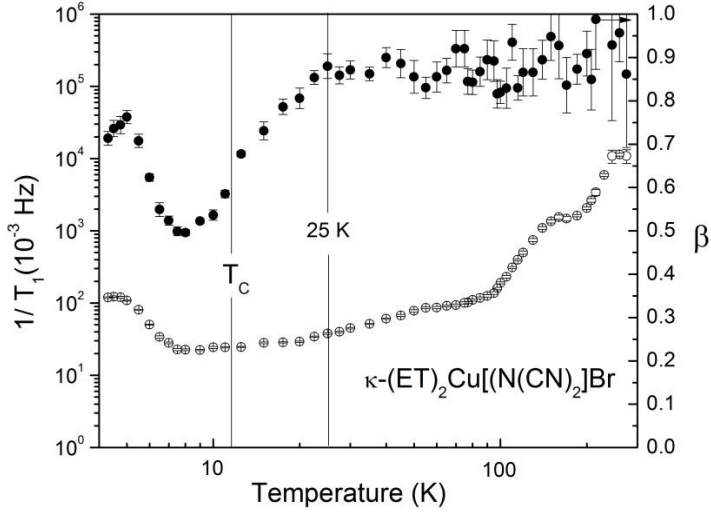

**Fig. 2.** Stretching exponent $\beta$ (filled circles) and $1/T_1$ (open circles) in κ-(ET)$_2$Cu[N(CN)$_2$]Br for $B_0 = 1$ T.

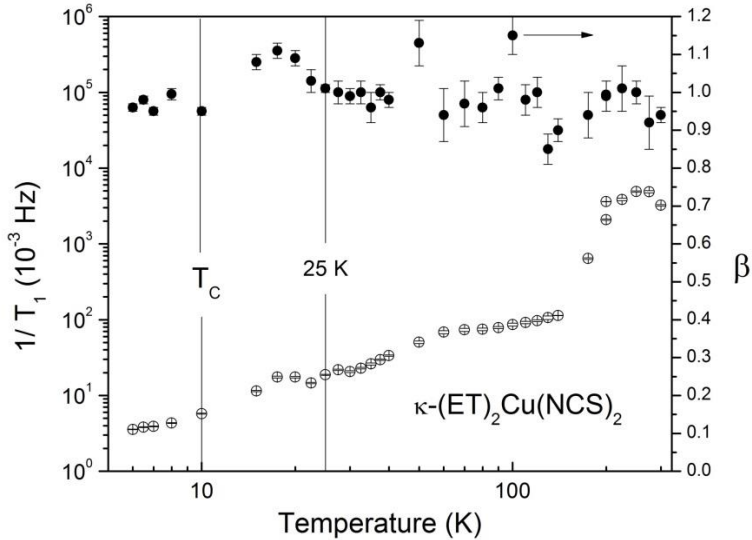

**Fig. 3.** Stretching exponent $\beta$ (filled circles) and $1/T_1$ (open circles) in κ-(ET)$_2$Cu(NCS)$_2$ for $B_0 = 1$ T.

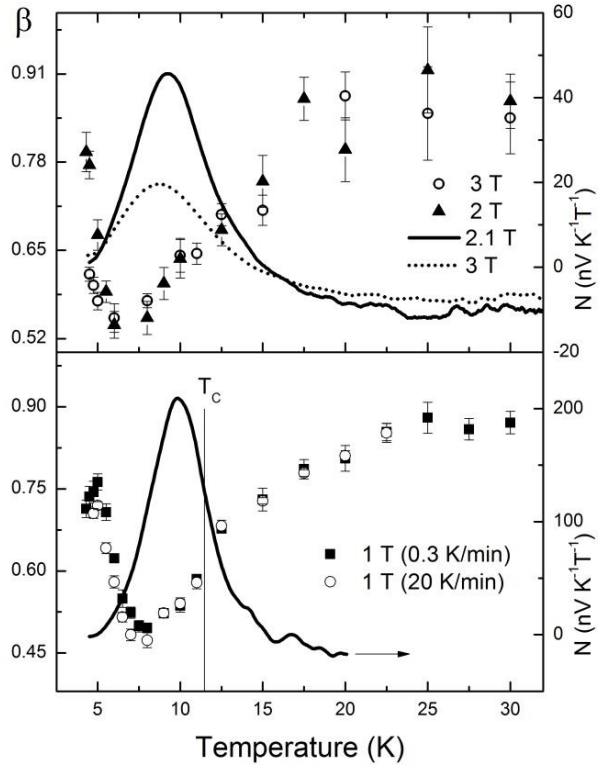

**Fig. 4.** (Top panel) Stretching exponent $\beta$ in $\kappa$-$(ET)_2Cu[N(CN)_2]Br$ for $B_0 = 2T$, $3T$ and Nernst coefficient for 2.1T, 3T. (Lower panel) $\beta$ for $B_0 = 1$ T with fast and slow cooling. Solid line shows Nernst coefficient for $B_0 = 1.1T$ [1].

In deuterated $\kappa$-$(ET)_2Cu[N(CN)_2]Br$, separation into metallic and antiferromagnetic regions was clearly demonstrated with both NMR[16] and scanning infrared spectroscopy [17]. The large internal field of the antiferromagnetic clusters resulted in $^{13}C$ NMR line splitting that developed below 30 K [16] and a broadening of the $^1H$ line below 15 K [44]. Our data show no line splitting or broadening near 25 K. $^{13}C$ NMR measurements on enriched samples, which are far more sensitive to changes in the conduction electron behavior, show a weak local minimum in the $^{13}C$ linewidth near 25 K but single exponential relaxation at all temperatures and no line splitting. If the stretched exponential behavior were due to phase separation one might expect it to depend upon cooling rate, but that is not the case. The lower panel of Fig. 4 shows essentially no difference in $\beta(T)$ for cooling rates of $dT/dt = 0.3$ K/min and $dT/dt = 20$ K/min. We therefore do not attribute the change at 25 K to phase separation.

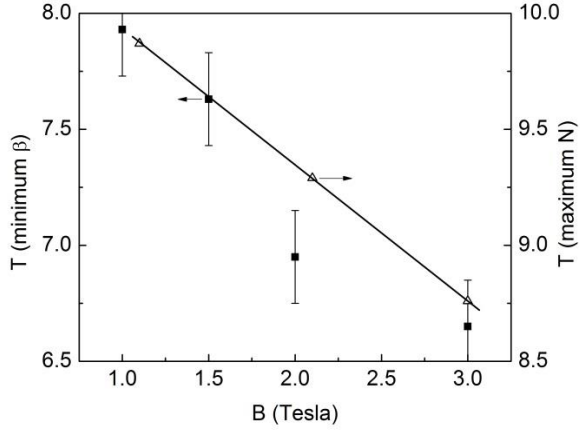

**Fig. 5.** Temperature of miniumum $\beta$ and maxiumum N versus field in $\kappa$-(ET)$_2$Cu[N(CN)$_2$]Br.

Fig. 4 shows the field dependence of $\beta$(T) for $B_0$ = 1, 2, and 3 T. As the field grows, the minimum value of $\beta$ grows larger and location of the minimum, T(minimum $\beta$), falls. The data is not precise enough to discern any systematic change above $T_C$. We have also plotted the Nernst coefficient, N, for $B_0$ = 1.1 T, 2.1 T and 3 T provided to us by the Oxford group [1]. The correspondence with $\beta$(T) is notable, each quantity showing an onset somewhere below 25 K, an extremum in the superconducting state and no discernible change at $T_C$. Fig. 5 shows the field dependence of T(minimum $\beta$) and T(maximum N). Both fiducial temperatures are determined by vortex motion, and track each other. Below $T_C$, a large N results from the flow of entropy-carrying vortices down a temperature gradient, in turn inducing a transverse electric field via the Josephson effect. The rapid increase of $\beta$(T) and $1/T_1$ below T(minimum $\beta$) is due to the increasing strength of magnetic field fluctuations from the vortex liquid, which scale as $1/\lambda^4$ where $\lambda$ is the penetration depth [45,46].

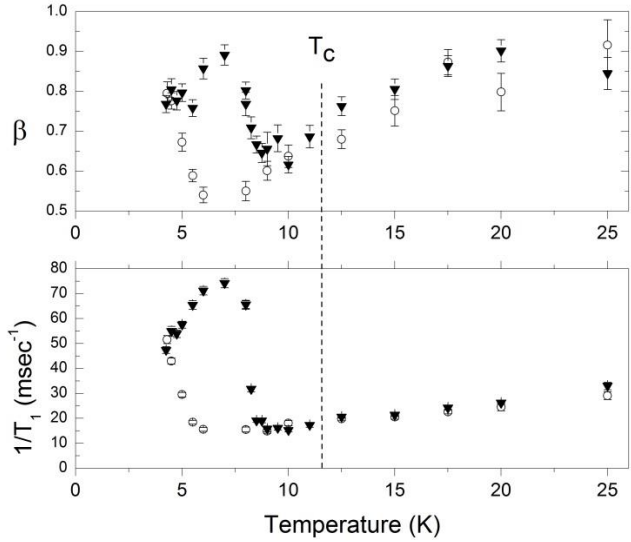

**Fig. 6.** (Top) Stretching exponent for field normal (circles) and parallel (triangles) to conducting planes. (Bottom) $1/T_1$ for the same two field orientations ($\kappa$-(ET)$_2$Cu[N(CN)$_2$]Br, $B_0$ = 2 T).

Vortex behavior in organic superconductors is highly anisotropic [10, 47,48]. If the stretched exponential relaxation were due to vortices we would expect this anisotropy to be reflected in the NMR. Fig. 6 shows $\beta$ and $1/T_1$ in $\kappa$-(ET)$_2$Cu[N(CN)$_2$]Br for two (nearly) orthogonal magnetic field orientations. Below $T_C$ both $\beta$ and $1/T_1$ vary strongly with orientation, with the peak in $\beta$ and $1/T_1$ occurring near the irreversibility temperature corresponding to the perpendicular field component [10-13,49]. However, above $T_C$, there is no significant orientation dependence. This would appear to rule out vortices as a source of spin relaxation above $T_C$. Normal state vortices may still be present and observable via transport measurements but they no longer generate field fluctuations on the time scale of nuclear spin precession. The stretched exponential relaxation below 25 K may signal the development of a new magnetic phase or density wave [9] that, itself, shows an anomalous Nernst signal.

Some time ago, Klutz et. al. [50] reported deviations from exponential recovery in collections of unoriented crystallites of κ-(ET)$_2$Cu(NCS)$_2$. The non-exponential relaxation began below $T = 25$ K and was attributed to nonuniform spin excitation due to a finite skin depth. Their model required a conductivity that *increases* with frequency. Our data do not support a skin depth picture. The RF field was oriented parallel to the conducting planes of each sample and for this orientation the skin depth (~ 3 mm) is larger than the κ-(ET)$_2$Cu[N(CN)$_2$]Br sample (1.2 x 0.7 x 0.6 mm). As shown in Fig. 4, the measurements at several different frequencies did not exhibit a systematic change in onset temperature as would be expected from a skin depth model. Moreover, the NMR nutation curves did not show the dependence upon spin rotation angle characteristic of skin-depth limiting [51]. Since the κ-(ET)$_2$Cu(NCS)$_2$ sample had lower resistivity and was larger (1.7 x 2.7 x 0.5 mm ), any finite skin depth effects would be enhanced over κ-(ET)$_2$Cu[N(CN)$_2$]Br. Instead, we found stretched exponential relaxation only in κ-(ET)$_2$Cu[N(CN)$_2$]Br, although the slope of $1/T_1T$ did deviate from a constant in the 10-30 K range (Fig.3). Within a chemical pressure scenario, κ-(ET)$_2$Cu(NCS)$_2$ resides further from the antiferromagnetic/superconductor phase boundary than κ-(ET)$_2$Cu[N(CN)$_2$]Br [3]. It is possible that strains incurred through packing of grains could lead to greater electronic inhomogeneity than obtained from a single crystal. The fact that the onset temperature occurred near 25 K is evidence for a magnetic energy scale common to the κ-(ET)$_2$X organics.

The behavior of $β(T)$ is reminiscent of a spin glass, for which simulations [52] and experiments [53] show that $β(T)$ falls monotonically beginning at a characteristic ordering temperature and approaches its asymptotic minimum near the glass transition. In κ-(ET)$_2$Cu[N(CN)$_2$]Br, a plausible magnetic ordering temperature would be $T_{Neel} = 27\,K$ observed in the closely related compound κ-(ET)$_2$Cu[N(CN)$_2$]Cl [18]. We looked for glass-like behavior by fixing the sample temperature at 15 K after rapid cooling and looking for an evolution of the NMR signal. Both the spectrum shape and location were unchanged over a period of 2500 minutes. This result does not, in itself, rule out a spin glass phase. Interestingly, recent μSR measurements of the local electronic spin susceptiblity in κ-(ET)$_2$Cu[N(CN)$_2$]Br exhibit a sharp peak near 15K and extending up to ~ 25K [54].

The data presented here have some similarities to underdoped $La_{2-x}Sr_xCuO_4$ which also exhibits a Nernst effect far above $T_C$ [5,6], stretched exponential $^{139}La$ NQR spin relaxation [19] and spin glass-like magnetic phases both inside and outside the superconducting dome [55]. The Nernst signal in Nd-doped $La_{2-x}Sr_xCuO_4$ shows behavior similar to $\kappa$-$(ET)_2Cu[N(CN)_2]Br$ when charge stripes are present but is featureless when they are absent [9]. Stripes have been proposed for the $\kappa$-$(ET)_2X$ materials [56] but these compounds reside at half-filling and appear to have conventional quasiparticles, casting doubt on a density wave reconstruction of the Fermi surface at 25 K. However, quantum oscillation experiments in $\kappa$-$(ET)_2Cu[N(CN)_2]Br$ require high fields (30 T) where magnetic breakdown is apparently dominant [57]. For some probes, breakdown may obscure Fermi surface modifications from a density wave with a small energy gap while the Nernst coefficient may be particularly sensitive to these changes. Large changes in the NMR spectrum do indeed take place in $\kappa$-$(ET)_2Cu[N(CN)_2]Br$ but at much higher temperatures [34,58]. Central $^{13}C$ nuclei, whose hyperfine coupling to conduction electrons is far larger than that of protons, undergo an abrupt six fold increase in linewidth near the 160 K peak in $1/T_1$ shown in Fig. 1 [34]. The proton NMR linewidth shows a somewhat less pronounced step. This linewidth transition might be a candidate for the onset of a density wave [59] or possibly a weak structural transition [58] but there is no evidence for stretched exponential relaxation anywhere near this region for either protons or $^{13}C$. Finally, Knight shift anomalies in $La_{2-x}Sr_xCuO_4$ [60] show evidence for a two component electronic fluid [61], similar to the situation in heavy fermion compounds [62, 63]. The appearance near 25 K of a magnetic phase in $\kappa$-$(ET)_2Cu[N(CN)_2]Br$ may be a manifestation of the same phenomenon.

In conclusion, the stretched exponential behavior observed in $\kappa$-$(ET)_2Cu[N(CN)_2]Br$ is evidence for electronic inhomogeneity that develops below 25 K and which coexists with superconductivity. The effect is absent in the closely related superconductor $\kappa$-$(ET)_2Cu(NCS)_2$ and this difference is also reflected in the appearance of a large normal state Nernst signal in roughly the same temperature range, but only in $\kappa$-$(ET)_2Cu[N(CN)_2]Br$. The onset temperature of 25 K reflects a magnetic energy scale that manifests itself throughout the $\kappa$-$(ET)_2X$ family of materials. Similar behavior observed in $La_{2-x}Sr_xCuO_4$ indicates that the normal state Nersnt signal as well as the coexistence of unconventional superconductivity and disordered magnetism are features common to the quasi-two dimensional organics and the copper oxides.

We wish to acknowledge conversations with N.P. Ong, E. Fradkin, M. Norman, R. Prozorov and A. Carrington. We thank S. Blundell, A. Ardavan and M.-S. Nam for useful comments and for providing us with their Nernst coefficient data. This work was supported by the National Science Foundation, grant # DMR-10-05708. Work at Argonne was supported by UChicago Argonne, LLC, Operator of Argonne